# Benchmark Tests of Slow Light in Saturable Absorbers

**A C Selden**

Department of Physics University of Zimbabwe

MP 167 Mount Pleasant HARARE Zimbabwe

e-mail address *adrian_selden@yahoo.com*

ABSTRACT

A series of benchmark tests of slow light in saturable absorbers is proposed. Stage I concerns experimental tests of saturable absorption, which can mimic slow light in saturable media. Stage II outlines the more demanding requirements for practical observation of spectral hole-burning in the absorption line, which is responsible for the reduction in group velocity.

**Introduction**

The theoretical equivalence of the 'slow light' and saturable absorption interpretations of signal transmission by saturable absorbers is well established [1, 2]. This equivalence can be traced to the perturbation treatment of wave interactions in saturable absorbers, in which the hole-burning and saturation terms are of equal magnitude, the former having a Lorentzian frequency profile, the latter being independent of frequency [3]. However, despite both qualitative and quantitative analysis of slow light experiments demonstrating their practical equivalence with saturable absorption [1, 2, 4], there is as yet no consensus concerning their correct interpretation i.e. whether they truly represent 'slow light', or arise from the delayed response of the absorbing transition. The phase shift and modulation gain of the transmitted signal – together with hole burning and power broadening in the modulation frequency spectrum – observed in single-beam CPO experiments [5-9], can be interpreted as either 'slow light' or the delayed response of the saturable absorber, without distinction [10]. Only when independent pump and probe beams are employed is there any practical means of distinguishing the two phenomena. When this is done, the probe can be tuned to scan the homogeneously broadened absorption profile and reveal the coherent hole created within it by the beating of pump and probe [11]. It is this which is responsible for the reduction in group velocity. The same method can also be used to test whether hole burning is actually necessary to produce the observed signal delay. A recent experiment on erbium doped optical fibre (EDF) provides a clear counter-example [12] in showing that the modulation phase may be either delayed or advanced by simply changing the relative phase of the modulated pump and probe for two counter-propagating beams at widely separated wavelengths within the broad absorption band. These observations are consistent with saturable absorption theory, but difficult to explain on the basis of slow light, given the enormous beat frequencies involved. The simple explanation offered, that the pump saturates the homogeneously broadened absorption band as a whole, the modulated probe undergoing a phase shift (delay or advance) determined by saturable absorption theory [12], seems the obvious choice in this case. Given the conflicting interpretations of single beam CPO experiments in saturable absorbers, some practical pump and probe tests are presented here for a range of saturable media, in the hope of stimulating a more critical approach to the investigation of slow light.

## Proposed experiments

### Stage I – saturable absorption

There seems no simple way of validating the original CPO 'slow light' experiments in ruby, alexandrite and bacteriorhodopsin (bR) [5, 6, 7], because the frequency width of the coherent hole for slow light in these media is a few tens of Hz (~0.1 Hz for bR film [7]) – too narrow to observe directly – while the pump sources used have spectral widths in the GHz range. In the absence of independent evidence, these experiments could be said to simulate slow light rather than demonstrate it, the results being indistinguishable from saturable absorption. They could be repeated using the same pump source (Ar ion or Ar-Kr laser) and an independent probe e.g. modulated diode laser, to test this [12]. As a refinement an etalon can be incorporated in the pump laser cavity to limit its frequency range by ensuring single longitudinal mode (SLM) operation. Potential combinations of saturable media and pump and probe sources for performing the initial benchmark tests are presented in Table I. In general, these represent minimal changes to the original CPO experiments, and are intended to test the saturable absorption model under virtually identical conditions. Ruby, alexandrite and bR can be tested by the method previously applied to erbium fibre, using an independent probe (diode laser) operating at a different wavelength to the modulated pump source [12]. 'Fast light' phenomena arising from reverse saturable absorption (RSA) in alexandrite [6] and fullerene [13] could be similarly tested.

### Pump source coherence

A further issue that could be addressed is the coherence of the pump source, usually an argon ion laser or a diode laser. In the former, the high spectral brightness and GHz bandwidth constitute an intense, partially coherent light source subject to rapid frequency fluctuations. The phase coherence of both pump and probe is presumably a necessary condition for coherent hole-burning. However, coherence is not necessary for saturating the absorption, as exemplified by the flashlamp pumped ruby laser [14]. This suggests a further test, using an incoherent source – such as an arc lamp – focused on the sidewall of the sample e.g. ruby, to partially saturate the absorption, and a coherent axial probe e.g. diode laser, to determine the phase shift and modulation gain, without creating a coherent hole.

## Linewidths of pump and probe beams

The original CPO slow light experiments utilised broadband pump sources of GHz linewidth modulated at audio frequencies in attempting to generate coherent holes of frequency width comparable to the inverse relaxation time, a few tens of Hz in ruby [5] and Er-fibre [8]. A benchmark test of slow light calls for both pump and probe sources to meet this restriction on frequency width, the probe frequency being precision tuned to sweep across the locked pump frequency, to scan the coherent hole created in the absorption band, an extreme combination of requirements, practically unattainable in these media. However, if we consider the beating of a single frequency probe with a broadband pump, we see that the population can only respond to beat frequencies on the order of the inverse relaxation time [15] and the delayed response of the medium itself meets the narrow frequency requirement. Secondly, when the pump frequency is subject to rapid fluctuations, its power spectrum has a narrow central peak (analogous to motional narrowing [16]) and can be effectively monochromatic [17]

## Stage II – spectral hole-burning experiments

The recent development of a <1 kHz linewidth, injection locked Ti:sapphire laser [18] enables some crucial slow light experiments to be conducted in saturable absorbers with sub-ms lifetimes, the pump source spectrum being of comparable width to the coherent hole. This is a minimum condition for performing single beam experiments with an amplitude modulated pump, but is not sufficient to validate slow light; a tunable probe of similar frequency width will be required for observing the coherent hole. Tunable diode lasers have been used in CPO slow light experiments in semi-conductor quantum well and quantum dot structures to scan the coherent hole in the optical spectrum [11, 19]. The relaxation times of semi-conductors lie in the ns range for interband transitions, enabling slow light experiments to be conducted with GHz width pump and probe beams. The phase shifts of the frequency-shifted sidebands can be detected independently via optical filtering [20]. The <1 kHz linewidth Ti:sapphire laser in conjunction with a narrow line dye laser (~200 Hz) [21] would be suitable for probing the anti-hole in alexandrite, with 260 μs lifetime for reverse saturable absorption [6, 22] The frequency doubled output of the Ti:sapphire would be matched to the absorption band of alexandrite. The same combination could also be used for probing the anti-hole in fullerene, responsible for 'fast light' [13]. However, a lower power frequency stabilised laser e.g. narrow line diode, could be used in place of the Ti:sapphire source because of the large absorption cross-section and relatively low saturation intensity of fullerene (Table II). Hole-burning experiments in saturable media with relaxation times of several ms e.g. ruby, Er-fibre, require light sources with linewidths of just a few tens of Hz, at least for the probe. Here we may look to the optical frequency comb, with component linewidths $\Delta\nu<40$ Hz [23], and lasers locked to ultra-stable reference cavities [24]. The 100 kHz linewidth required for probing the coherent hole in $Cr^{4+}$:YAG, commonly used as a Q-switch in solid state laser systems[25], could be met by an extended cavity diode laser [26], but a high intensity pump source will be required to saturate the absorption e.g. a Ti:sapphire laser. $Cr^{4+}$:YAG also undergoes excited state absorption (ESA), which could be probed with this experimental setup. Spectral hole-burning observed in early experiments on intra-cavity dye cells used as passive laser Q-switches [27] could be studied using a repetitively Q-switched Ti:sapphire laser as the pump source, combined with a tunable, pulsed diode laser probe. The relevant parameters of a broad range of saturable media are presented in Table II, together with pump and probe sources for conducting the hole-burning experiments. Those used for directly probing the coherent hole in semiconductor nanostructures, namely the original experiment on a GaAs-AlGaAs quantum well waveguide [11] and InAs-GaAs quantum dot [19], are included for reference.

**Conclusions**

Given the current ambiguity in the interpretation of CPO slow light experiments in saturable absorbers – and a recent counter-demonstration in erbium fibre [12] – more definitive experiments on signal transmission in saturable media are needed. A series of benchmark tests proposed here may help resolve this ambiguity for a wide range of saturable absorbers.

**Acknowledgment**

I am indebted to Valerii Zapasskii for first bringing the ambiguity of slow light in saturable absorcbers to my attention and for his valued comments on this subject. I also wish to thank Bruno Macke for sharing his insights on slow light and kindly providing reference material related to motional narrowing.

**Table I**

| Sample | $\lambda_s$ | $I_s$ | $\tau_s$ | $\Delta\nu_h$* | Pump | Probe | Refs |
|---|---|---|---|---|---|---|---|
| Ruby | 514.5 nm | 1.5 $kWcm^{-2}$ | 4.45 ms | 36 Hz | Ar-ion | DL$^\$$ | 5 |
| Alexandrite | 488 nm<br>457 nm | ~ $kWcm^{-2}$ | 19 ms<br>260 μs | 8.5 Hz<br>612 Hz$^\#$ | Ar-ion | DL | 6 |
| Fullerene | 514.5 nm | 1 $kWcm^{-2}$ | 143 μs$^‡$ | 1.1 kHz$^\#$ | Ar-ion | DL | 13, ‡28 |
| EDF | 1536.2 nm | 3.4 $kWcm^{-2}$ | 10.5 ms | 10 Hz | DFB DL | DL | 8, 12 |
| bR film | 568.2 nm | ~ $mWcm^{-2}$ | 1.43 s | 0.1 Hz | Ar-Kr | DL | 7 |
| bR aq. | 568.2 nm<br>647.1 nm | ~ 3 $Wcm^{-2}$ | ~5 ms | ~50 Hz<br>~15 Hz$^\#$ | Ar-Kr | DL | 29 |

*HWHM
#anti-hole

$diode laser

## Table II

| Sample | $\lambda_s$ | $I_s$ | $\sigma_s$ cm$^2$ | $\tau_s$ | $\Delta\nu_h$* | Pump | Probe | Refs |
|---|---|---|---|---|---|---|---|---|
| Alexandrite | 476 nm | $^{\dagger}$180 kWcm$^{-2}$ | 0.9 x10$^{-20}$ | 260 μs | 612 Hz$^{\#}$ | Ti:sapphire$^{\$}$ | Dye laser$^{\mu}$ | 6, 22 |
| Fullerene | 450 nm | $^{\dagger}$1 kWcm$^{-2}$ | 3.1 x10$^{-18(c)}$ | 143 μs$^{(d)}$ | 1.1 kHz$^{\#}$ | Ti:sapphire$^{\$}$ | Dye laser$^{\mu}$ | $^{(d)}$28, $^{(c)}$30 |
| $Cr^{4}$:YAG | 1 μm | $^{\dagger}$13 kWcm$^{-2}$ | 4.3 x10$^{-18(e)}$ | 3.5 μs$^{(f)}$ | 50 kHz | Ti:sapphire$^{\ddagger}$ | TDL$^{¶}$ | $^{(e)}$31, $^{(f)}$25 |
| CuPc-PMMA | 700-800 nm | $^{\dagger}$9 MWcm$^{-2}$ | 2 x10$^{-18(g)}$ | 15 ns$^{(g)}$ | 10 MHz | Ti:sapphire$^{+}$ | TDL$^{+}$ | $^{(g)}$32 |
| AlCl-Pc | 700-750 nm | $^{\dagger}$35 MWcm$^{-2}$ | 1.5 x10$^{-18(h)}$ | 5 ns$^{(i)}$ | 30 MHz | Ti:sapphire$^{+}$ | TDL$^{+}$ | $^{(h)}$33, $^{(i)}$34 |
| DTTCI | 700-800 nm | ~100 MWcm$^{-2}$ | | 1.34 ns$^{(j)}$ | 120 MHz | Ti:sapphire$^{+}$ | TDL$^{+}$ | $^{(j)}$35 |
| GaAs-AlGaAs (QW) | 820 nm | 1.6 kWcm$^{-2\wedge}$<br>2.5 kWcm$^{-2\wedge\wedge}$ | | 0.33 ns$^{\wedge}$<br>0.51 ns$^{\wedge\wedge}$ | 500 MHz | Ti:sapphire$^{§}$ | TDL | 11 |
| InAs-GaAs (QD) | 1291 nm | | | | ~ 2 GHz | DFB DL | TDL | 19 |

$^{\dagger}$calc   $^{\wedge}\Delta\theta = 5$ deg   $^{\wedge\wedge}\Delta\theta = 2.5$ deg

*HWHM
$^{\#}$anti-hole

$^{\$}\Delta\nu \approx 1$ kHz freq doubled
$^{\ddagger}\Delta\nu \approx 100$ kHz $^{\mu}\Delta\nu \approx 200$ Hz [21]
$^{¶}$tunable DL $^{+}$rep pulsed $^{§}$single mode